\documentclass[twocolumn,showpacs,preprintnumbers]{revtex4}

\usepackage{graphicx}
\usepackage{dcolumn}
\usepackage{bbm}
\usepackage{amssymb}
\usepackage{amsmath}
\usepackage{epsfig}

\def\lapp{\mathrel{\rlap{\raise.5ex\hbox{$<$}}
                    {\lower.5ex\hbox{$\sim$}}}}
\def\gapp{\mathrel{\rlap{\raise.5ex\hbox{$>$}}
                    {\lower.5ex\hbox{$\sim$}}}}

\newcommand{\bb}{\mathbbm}
\newcommand{\thb}{{\theta_b}}
\newcommand{\beq}{\begin{equation}}
\newcommand{\eeq}{\end{equation}}
\newcommand{\beqa}{\begin{eqnarray}}
\newcommand{\eeqa}{\end{eqnarray}}

\def\bc {\begin{center}}
\def\ec {\end{center}}
\def\be {\begin{equation}}
\def\bea {\begin{eqnarray}}
\def\ee {\end{equation}}
\def\eea {\end{eqnarray}}

\begin{document}

\title { 
Probing $CPT$ violation in neutrino oscillation: A three flavor analysis
}

\author
{\sf Abhijit Samanta\footnote{Email: abhijit@hri.res.in}}
\affiliation
{\em Harish-Chandra Research Institute,
Chhatnag Road,
Jhusi,
Allahabad 211 019,
India\footnote{Present address:
 Ramakrishna Mission Vivekananda University, Belur Math, Howrah 711 202,
India}}

\date{\today}

\begin{abstract}
\noindent
We have  studied $CPT$ violation in neutrino oscillation considering 
three flavor framework with matter effect. 
We have constructed  a new way to find the oscillation probability 
incorporating  $CPT$ violating terms without any approximation. 
Then  $CPT$ violation with 
atmospheric neutrinos for a magnetized iron calorimeter detector
considering the muons (directly measurable with high resolution) 
of the charge current events has been studied for zero and nonzero 
$\theta_{13}$ values. It is found that a potential bound of
$\delta b_{32} \lapp 6\times 10^{-24}$ GeV at 99\% CL can be 
obtained with  1 Mton.year exposure of this detector; and
unlike neutrino beam experiments, there is no  possibility 
to generate `fake' $CPT$ violation due to matter effect with atmospheric neutrinos.
The advantages of atmospheric neutrinos to discriminate 
$CPT$ violation from $CP$ violation and
nonstandard interactions are also discussed.
\end{abstract}


\maketitle
\section{Introduction}

The  invariance of the product of charge conjugation ($C$), parity ($P$) and time reversal ($T$)
 $CPT$ is one of the most fundamental symmetries in physics and it is indeed  related intimately with 
Lorentz invariance. 
The $CPT$ theorem \cite{cpttheorem} states that any local quantum field 
theory (QFT) which is Lorentz invariant and has a Hermitian Hamiltonian must have
$CPT$ symmetry.  
The general theoretical proof of $CPT$ invariance in particle 
physics along with accurate experimental tests of $CPT$ violation ($CPTV$)
provides  an attractive candidate
signature for non particle physics such as string theory \cite{kp1,kp2}.
In particular, the assumptions needed to prove $CPT$ theorem are invalid for strings
which are extended objects. Moreover, since the critical string dimensionality is greater
than four, it is possible that higher dimensional Lorentz invariance breaking would be incorporated
in a realistic model. 
O.W. Greenberg has shown that $CPT$ violation implies violation of Lorentz invariance,
but, CPT invariance is not sufficient for out-of-cone Lorentz invariance
\cite{Greenberg:2002uu}.

Consequences of $CP$, $T$ and $CPT$ violation in neutrino oscillation 
have been discussed in \cite{CPTV}. Briefly, for 
$\nu_\alpha\to\nu_\beta$ flavor oscillation probabilities
$P_{\alpha\beta}$ at a distance $L$ from the source is the
following. If
\begin{equation}
P_{\alpha\beta}(L) \neq P_{\bar\alpha\bar\beta}(L) \,,
\qquad \beta \ne \alpha \,,
\end{equation}
then $CP$ is not conserved.
If
\begin{equation}
P_{\alpha\beta}(L) \neq P_{\beta\alpha}(L) \,,
\qquad \beta \ne \alpha \,,
\end{equation}
then $T$-invariance is violated.
If 
\begin{eqnarray}
P_{\alpha\beta}(L) &\neq& P_{\bar\beta\bar\alpha}(L)\,,
\qquad \beta \ne \alpha \,,
\\
\noalign{\hbox{or}}
P_{\alpha\alpha}(L) &\neq& P_{\bar\alpha\bar\alpha}(L) \,,
\end{eqnarray}
then $CPT$ is violated.
Also it has been shown that matter effects give rise to apparent  
$CP$ {\it and} $CPT$ violation even if the mass matrix is $CP$ conserving.


At the experimentally accessible energies, the signals for Lorentz
and $CPTV$  have been described by a class of extensions of the Standard
Model (SME) \cite{others,Colladay:1996iz,Coleman:1998ti,CKcomb}. 
The low energy probes of these new physics
have been explored in many experiments with current technologies \cite{lit}.
A summary on measured and derived values of the coefficients for Lorentz
and $CPTV$ in SMEs have been nicely tabulated in \cite{Kostelecky:2008ts} 
for matter, photon and gravity sectors. 

The SME framework predicts several unusual phenomena in neutrino oscillation, 
among which are neutrino-antineutrino oscillation, directional dependence,
unconventional energy behavior \cite{Kostelecky:2003cr}.  These effects have been explored 
with recent experimental data at MINOS \cite{:2008ij} and at LSND 
\cite{Auerbach:2005tq} 
and new bounds have been obtained. The effect of perturbative Lorentz and $CPTV$ 
on neutrino oscillation has also been studied in \cite{Diaz:2009qk} with leading order
corrections arising from renormalizable operators including the above 
phenomena. More consequences of different 
SMEs of Lorentz and $CPTV$ in neutrino oscillation can be found in 
\cite{Kostelecky:2003xn,Katori:2006mz,Kostelecky:2004hg} for atmospheric
solar and baseline experiments as well as in \cite{Bustamante:2010nq,Bhattacharya:2010xj} 
for ultra-high energy neutrino experiments.

With an effective-theory approach, a plausible spontaneous $CPT$ violating 
minimal extension of standard model that are also Lorentz invariance violating (LV) 
has been done in \cite{Colladay:1996iz} for Dirac particles and in \cite{Coleman:1998ti}
for Majorana particles. Our analysis applies for both Dirac or Majorana neutrinos.

In these theories the 
Lagrangian for a fermion to the lowest order in the high scale can  be written as

\bea
{\cal L} = i \bar{\psi} \partial_\mu \gamma^\mu \psi 
-m \bar{\psi} \psi            
- A_\mu \bar{\psi} \gamma^\mu \psi 
- B_\mu \bar{\psi} \gamma_5 \gamma^\mu \psi  \; ,
\label{L-psi}
\eea
where, $A_\mu$ and $B_\mu$ are real numbers.
The terms containing $A_\mu$ and $B_\mu$ are clearly 
Lorentz invariance violating. 
The effective 
contribution from these terms to the neutrino Lagrangian 
can be parametrized as 
\bea
{\cal L}_\nu^{CPTV} =  
\bar{\nu}_L^\alpha \, b^{\alpha \beta}_\mu \, 
\gamma^\mu \, \nu_L^{\beta} \; ;
\label{L-nu}
\eea
where, $b_\mu$ are four Hermitian $3\times 3$
matrices corresponding to the four Dirac indices $\mu$ and
$\alpha, \beta$ are flavor indices.
The effective Hamiltonian for 
ultra-relativistic neutrinos with a definite momentum $p$ is then
\bea 
{\bb H} \equiv \frac{{\bb M} {\bb M}^\dagger}{2 p} + {\bb b} \; , 
\label{eff-H}
\eea
where 
${\bb M}$ is the neutrino mass matrix 
and ${\bb b} \equiv b_0$ 
in the 
$CPT$ conserving limit. 
If we choose a preferred frame in which the Cosmic Microwave Background Radiation (CMBR) 
is isotropic, then the rotational invariance implies no directional
dependence for ${\bb b}$ \cite{Coleman:1998ti}.

In presence of LV, one can obtain
the same effective Hamiltonian by considering 
a modified dispersion relation for neutrinos, 
$E^2 = F(p,m)$.
Using the rotational
invariance in the CMBR frame and demanding Lorentz invariance 
at low energy, this dispersion relation can be written as
\cite{Mattingly:2005re}

\bea E^2 = m^2 + p^2 + E_{Pl} f^{(1)} |p| + f^{(2)} |p|^2 + \frac{f^{(3)}}{E_{Pl}} |p|^3 \cdots, \eea

where, $f^{(n)}$'s are dimensionless quantities. $E_{Pl}$ is the Planck scale of energy where
the Lorentz invariance is expected to be broken.   For ultra relativistic neutrinos with fixed 
momentum, the above dispersion relation becomes

\bea  E = p + \frac{m^2}{2p} + b  \cdots, \eea

with $b=E_{Pl} f^{(1)} /2$ as the leading $CPT$ violating contribution. For three flavors, it leads 
to the same effective Hamiltonian as in Eq. \ref{eff-H}.

The $CPTV$ in neutrino oscillation was first proposed for two flavor case in 
\cite{Coleman:1998ti}. The typical frequency of neutrino oscillation is 
$\Delta m^2 / (2 E)$. For atmospheric and long baseline neutrinos, 
it can be as small as $10^{-22}$ GeV. If the accuracy of the oscillation 
frequency is 10\%, then one can naively estimate the $CPTV$ parameter to the 
order of $10^{-23}$ GeV.

The $CPTV$ in two flavor formalism has been studied for the future atmospheric 
neutrino experiment
at a magnetized iron calorimeter (ICAL) detector \cite{Datta:2003dg} and for long baseline
(735 km) experiment with a typical neutrino factory \cite{Barger:2000iv}. In  \cite{Barger:2000iv}, 
it has been
shown that  the interference between the $CPT$ violating interaction and $CPT-$even 
mass terms in the Lagrangian can lead to a resonant enhancement of the 
oscillation amplitude and this may lead to `fake' $CPT$ violation.
In \cite{Bilenky:2001ka}, it has been shown that 
for hierarchical mass spectrum    
the upper bound of the neutrino antineutrino mass difference can be achieved in a 
neutrino factory is $|m_3-\bar m_3|\lapp 1.9 \times 10^{-4}$ eV.

The LSND result \cite{LSND} when combined with the solar and atmospheric neutrino
observations indicated
three distinct neutrino mass squared differences.
Then it was proposed that the CPT violating effects may be
large enough to make the neutrino and antineutrino
spectra significantly different
\cite{LSND-Smirnov,Yanagida}.
However, this  fact was found not to be viable when
combined with other neutrino experiments \cite{gouvea-cpt},
and the subsequent data of oscillations corresponding to
$\Delta m^2_\odot$ in antineutrinos at KamLAND \cite{kamland}
ruled it out.
The CPT violation is not required to explain any neutrino oscillation
data if the LSND results are ignored in the light of the
negative results of MiniBooNE \cite{MiniBooNE} that
explore the same parameter space.
However, the current uncertainties in the measurements of
$\Delta m_{21}^2$ and $\Delta m_{32}^2$ \cite{Fogli:2008ig}, allow 
the possibility to study CPT violating effects
in neutrino oscillations, which may be observed or constrained
at the future high precision neutrino oscillation experiments.

From the  present experimental data, in Ref. \cite{Bahcall:2002ia}, it is shown that
$\delta b \lapp 1.6\times 10^{-21}$ GeV with solar and KamLAND data,
and,  in Ref. \cite{GonzalezGarcia:2004wg}, $\delta b \lapp 5\times 10^{-23}$ GeV 
with full atmospheric and K2K data for two flavor analysis.

The three flavor analysis of $CPTV$  is mainly needed to account properly the 
matter effect and $CP$ violating phases. 
If ${\bb b}$ is 
not diagonal in mass basis,  one needs to introduce  
an unitary matrix  $\mathbbm{U}_b$ to diagonalize it. 
The matrix $\mathbbm{U}_b$ then needs three angles ${\thb}_{12}$, 
${\thb}_{23}$, $\thb_{13}$ and six phases for a complete
parametrization and it can be written as 
\bea
\mathbbm{U}_b(\{\thb_{ij}\}; \{\phi_{bi}\};\{\alpha_{bi}\};\delta_{b}) &&\nonumber  \\
 =  {\rm diag}(1, e^{i\phi_{b2}}, e^{i \phi_{b3}}) \cdot
\mathbbm{U}_{CKM}(\{ \thb_{ij}\}; \delta_b) \cdot \nonumber && \\
  \phantom{space}
{\rm diag}(e^{i \alpha_{b1}}, e^{i \alpha_{b2}}, e^{i \alpha_{b3}}) \; .
\eea
where, $\alpha_{b1}$, $\alpha_{b2}$, $\alpha_{b3}$ 
are the Majorana phases and will not 
have any contribution. To diagonalize the effective Hamiltonian, the mixing matrix
appears  in term of a total of 
six mixing angles $(\theta_{12}, \theta_{23}, \theta_{13},
\thb_{12}, \thb_{23}, \thb_{13})$ and four phases
$(\delta_{cp}, \delta_b, \phi_{b2}, \phi_{b3})$.
This
makes the marginalization of the data practically impossible with normal CPUs. 

In \cite{Dighe:2008bu}, the authors have analyzed $CPTV$ for three flavor case and 
have treated the effect of the CPT violating term as a perturbation parametrized
by a dimensionless auxiliary parameter $\epsilon \equiv 0.1$.
Finally, they identify the combinations of CPT
violating parameters that contribute to the probabilities
to leading order in $\epsilon$ and compare the signals in different channels
to estimate the extent to which these CPT violating combinations
can be constrained or identified
in future long baseline experiments.

 In this paper, we assume
 $\bb b$ matrix to be diagonal in mass basis. 
Then, there arises only two parameters $\delta b_{21}=b_2-b_1$ and $\delta b_{32}=b_3-b_2$. 
we have introduced a new
method following Ref. \cite{Barger:1980tf}, where diagonal  $\bb b$ matrix 
directly comes into the picture.
So, in this
formalism, there is no need to transfer $\bb b$ to flavor basis. Here, we have also 
assumed a preferred frame in which the CMBR
is isotropic, then the rotational invariance implies no directional
dependence for ${\bb b}$ \cite{Coleman:1998ti}.

Then with this three flavor formalism, we  have studied $CPTV$ with atmospheric 
neutrinos for a magnetized ICAL detector proposed at the India-based Neutrino
Observatory (INO) \cite{ino} considering the muons (directly measurable 
quantities at ICAL). In this type of detector, one can separate neutrinos
and antineutrinos due to the magnetic field.

\section{Neutrino oscillation in matter with $CPT$ violation}

We have  incorporated $CPTV$ interactions in a new way where two 
$CPTV$ terms come directly in the oscillation probability without any approximation. 
We derived this formalism from the original work in \cite{Barger:1980tf}.
We discuss it here for any number of generations.

The neutrino flavor eigenstates $|\nu_\alpha (\alpha = e, \mu, \tau ...)$ and mass
eigenstates $|\nu_i (i = 1, 2, 3 ...)$ at time $t=0$ are connected by a unitary
transformation,

\bea |\nu_\alpha \rangle = \sum_i U_{\alpha i} |\nu_i \rangle . \eea

Then for a relativistic neutrino beam with energy $E$, the standard amplitude $A$ and 
probability $P$ for $\nu_\alpha \rightarrow \nu_\beta$ transition after time $t$ 
in vacuum are

\bea A (\nu_\alpha \rightarrow \nu_\beta) = \sum_i U_{\alpha i} 
exp (-\frac{1}{2} i m_i^2 t /E) U^\dagger_{i\beta},\eea

\bea 
 P (\nu_\alpha \rightarrow \nu_\beta)=
|A (\nu_\alpha \rightarrow \nu_\beta) |^2;
\eea

where, $m_i$ are the mass eigen values and 
since neutrinos are ultra relativistic $t/E\equiv L/E$ in units $\hbar =c =1$. 

To treat neutrino oscillation in matter, let us introduce an arbitrary state vector 
in neutrino flavor space,
\bea
|\psi\rangle = \sum_i \psi_i(t) |\nu_i\rangle 
\eea

Now, at time $t=0$, consider the state as $\nu_{\alpha}$. Then $\psi_i(0) = U_{\alpha i}$
and the transition amplitude is 
\bea A (\nu_\alpha \rightarrow \nu_\beta)=\sum_i U^\dagger_{i\beta} \psi_i(t).\eea 

Then time evolution in presence of matter and $CPTV$ term becomes

\bea id\psi_j(t)/dt&=&\left [m_j^2/(2E)+ b_j\right ] \psi_j(t)\nonumber\\
&+&\sum_k\sqrt{2}GN_e U_{ej} U^\dagger_{ke}\psi_k(t) 
\nonumber\\
&\equiv& H_{kj}\psi_k(t)\label{e:h}\eea
assuming $\bb b$ to be diagonal in mass basis.
Here,  $N_e$ and  $G$ are the electron number density of the medium and Fermi constant, respectively.
This equation is for neutrinos and for antineutrinos, the sign of $\sqrt{2}GN_e$ and $b_i$ 
will be reversed and $U$ will be replaced by $U^*$.

The problem of propagation is therefore to diagonalize the matrix $H$ defined in Eq. 
\ref{e:h}. It can be diagonalized by defining a new basis states
$|\nu^\prime_i\rangle = V^\dagger_{i\alpha}|\nu_\alpha\rangle = V^\dagger_{i\alpha} U_{\alpha j}
|\nu_j\rangle$, where $V$ is the unitary matrix.   
If $M^2_i/(2E)$ are the set of eigen values of $H$, then the solution for a uniform medium
is given by replacing $m_i$ by $M_i$ and $U$ by $V$, respectively. 

However, one can find the solution  without explicitly finding $V$ in  the following way.
For $n$ generation, Eq.. \ref{e:h} has $n$ independent solution for the row
vectors $\psi_j(t)$. At time $t=0$, we choose the set of solutions 
$\psi_j^{(i)}~ (i=1, \cdots n)$ for the row vectors. These are pure mass eigen states.

\bea \psi_j^{(i)} (t=0)=\delta_{ij}\eea 

When these row vectors are assembled into a $n\times n$ matrix $X$ according to 

\bea X_{ij}(t)= \psi_j^{(i)}(t),\eea

then $X$ follows the matrix equation

\bea idX/dt=XH\eea
with boundary condition $X(t=0) =1.$ 
If $N_e$ is constant, an analytical solution is possible,

\bea X(t)=exp(-iHt).\eea

The row $i$ of $X$ matrix describes the state $|\nu_i\rangle$ which is a mass eigen state
at the starting point $(t=0)$ and column $j$ describes the amplitude for evolving into the 
mass eigen state $|\nu_j\rangle$ at time $t$.
Then the transition amplitude in matter is 

\bea A(\nu_\alpha  \rightarrow \nu_\beta) = \sum_{ij} U_{\alpha i} X_{ij} U^\dagger_{j\beta}\eea

where,
\bea X=\sum_k \left[\Pi_{j\ne k}\frac{(2EH-M_j^2 1)}{\delta M_{kj}^2} \right] 
exp\left ( -i\frac{M_k^2 L}{2E}\right )
\eea
with $\delta M_{kj}^2=M_k^2-M_j^2$.
If one subtracts $m_1^2/(2E)$ and $b_1$, which will not alter the oscillation probability,
there will appear $\Delta m_{j1}^2$ and $\delta b_{j1}=b_j-b_1$ in the final formula.

We have developed a numerical program considering the above formalism using 
Preliminary Reference Earth Model (PREM)  
\cite{Dziewonski:1981xy} for the density profile of the Earth to find $X$ and 
then the oscillation probabilities.
We have checked that when $\delta  b_{ji}=0$, the result matches exactly 
with the other methods of finding oscillation probabilities in matter.

\section{$CPT$ violation with atmospheric neutrinos}

\subsection{The chi-square analysis}

We have studied the atmospheric neutrino data  for the magnetized ICAL detector 
considering the muon energy and direction (directly measurable quantities)
of the events in context of CPT violation.
Due to relatively heavy mass of muon, it looses energy mostly via ionization and atomic 
excitation during its propagation through a medium. Since  ICAL is a tracking detector,
 it gives a clean track in the detector. The muon energy can be measured from the bending of the track in
magnetic field or from track length in case of fully contained event. The direction can be
measured from the tangent of the track at the vertex.
From the GEANT \cite{geant} simulation of ICAL detector, it is clear that 
the energy and
angular resolutions of the muons are very
high (4-10\% for energy and 4-12\% for zenith angle) and 
negligible compared to the resolutions obtained
from the kinematics of the scattering processes. In our analysis, we neglect the
effect arisen due to the properties of the detector.

A new method for migration from true neutrino
energy and zenith angle to muon energy and zenith angle has been introduced in \cite{Samanta:2006sj}
and subsequently used in \cite{Samanta:2008af,Samanta:2008ag,Samanta:2009hd,
Samanta:2009qw,Samanta:2010zh}. Here we used NUANCE-v3 \cite{Casper:2002sd}
for generating the events.
The addition of the hadron energy to the muon energy, which might improve
the reconstructed neutrino energy resolution, is not considered here for
conservative estimation of the sensitivity. It would be realistic in case
of GEANT-based studies since the number of hits produced by the
hadron shower   strongly depends on the thickness of iron layers.

The $\chi^2$ is calculated according to the Poisson probability distribution.         
The binning the data is made  in  2-dimensional grids
in the plane of $\log_{10} E$ - $L^{0.4}$. 
For each set of oscillation parameters, we integrate the oscillated
atmospheric neutrino flux folding the cross section, the exposure time, the target mass, the efficiency
and the two dimensional energy-angle correlated resolution functions to obtain the predicted data
for the $\chi^2$ analysis.
We use the charge current cross section of Nuance-v3 \cite{Casper:2002sd} and the
Honda flux  in 3-dimensional scheme \cite{Honda:2006qj}.
This method has been  introduced in \cite{Samanta:2006sj}, but the number of
bins and resolution functions have been optimized later 
\cite{Samanta:2008af,Samanta:2008ag,Samanta:2009hd,Samanta:2009qw}.
 The method for migration of number of events
from neutrino to muon energy and zenith angle bins, the number of bins, the systematic                    
uncertainties, and the cuts at the near horizons are described in \cite{Samanta:2008ag}.     
Both theoretical and experimental data for $\chi^2$ analysis have been generated in the 
same way by migrating number of events from neutrino to muon energy and zenith angle bins 
using the resolution functions, which has been used in our previous work \cite{Samanta:2009qw}.

We marginalize the $\chi^2$ over the oscillation parameters
$\Delta m_{32}^2,~\theta_{23}, ~\theta_{13}$, $\delta_{CP}$, $\delta b_{21}$ and $\delta b_{32}$
for both normal hierarchy and inverted hierarchy
with $\nu$s and $\bar\nu$s separately for a given set of input data.
Then we find the total $\chi^2$ as $\chi^2=\chi^2_\nu+\chi^2_{\bar\nu}$.

We have chosen the range of $\Delta m_{32}^2=2.0-3.0\times 10^{-3}$eV$^2$,
$\theta_{23}=37^\circ - 54^\circ,$ $\theta_{13}=0^\circ - 12.5^\circ$
and $\delta b_{21,32}=0-5\times 10^{-23}$.
$\Delta m_{21}^2=7.06-8.34\times 10^{-5}$eV$^2$ and $\theta_{12}=30.5^\circ - 40^\circ.$
%
However, the effect of $\Delta m_{21}^2$ comes in the
subleading order in the oscillation probability when $E\sim$ GeV and it is marginal.
%
We have set the inputs of $\Delta m_{32}^2=-2.5\times 10^{-3}$eV$^2$,
$\theta_{23}=45^\circ$  and $\delta_{CP}=0$.

\subsection{Result and discussions}
\begin{figure*}[htb]
\includegraphics[width=8.0cm,angle=0]{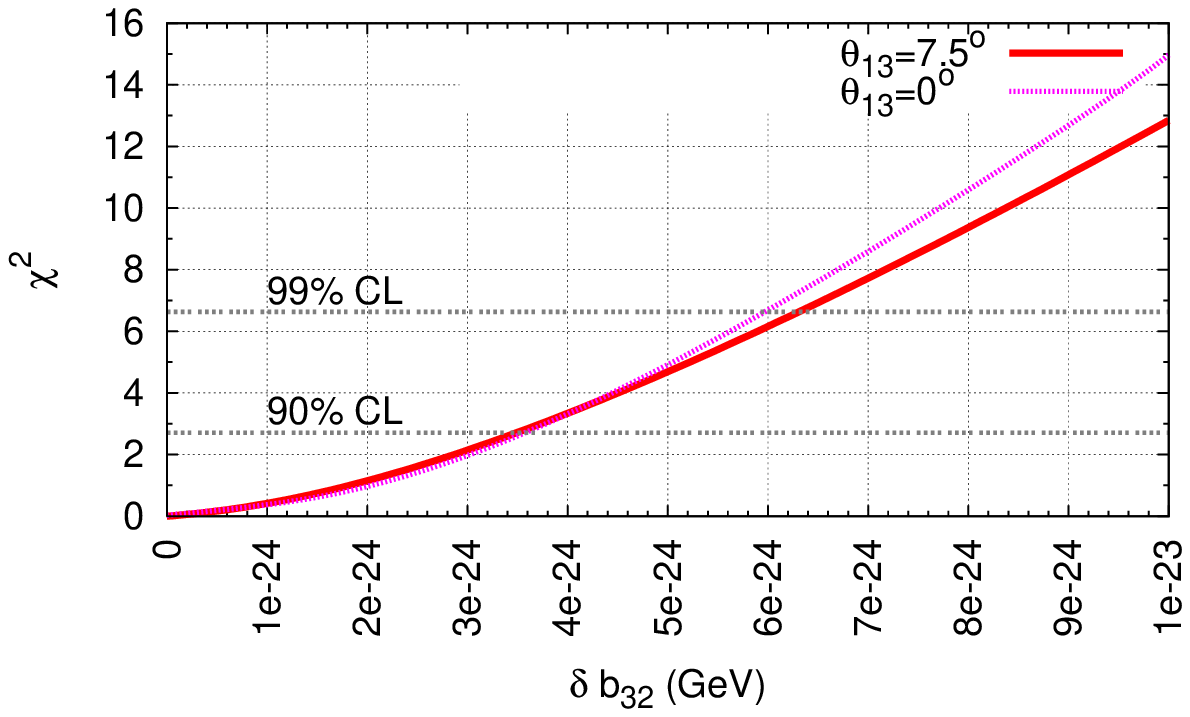}
\includegraphics[width=8.0cm,angle=0]{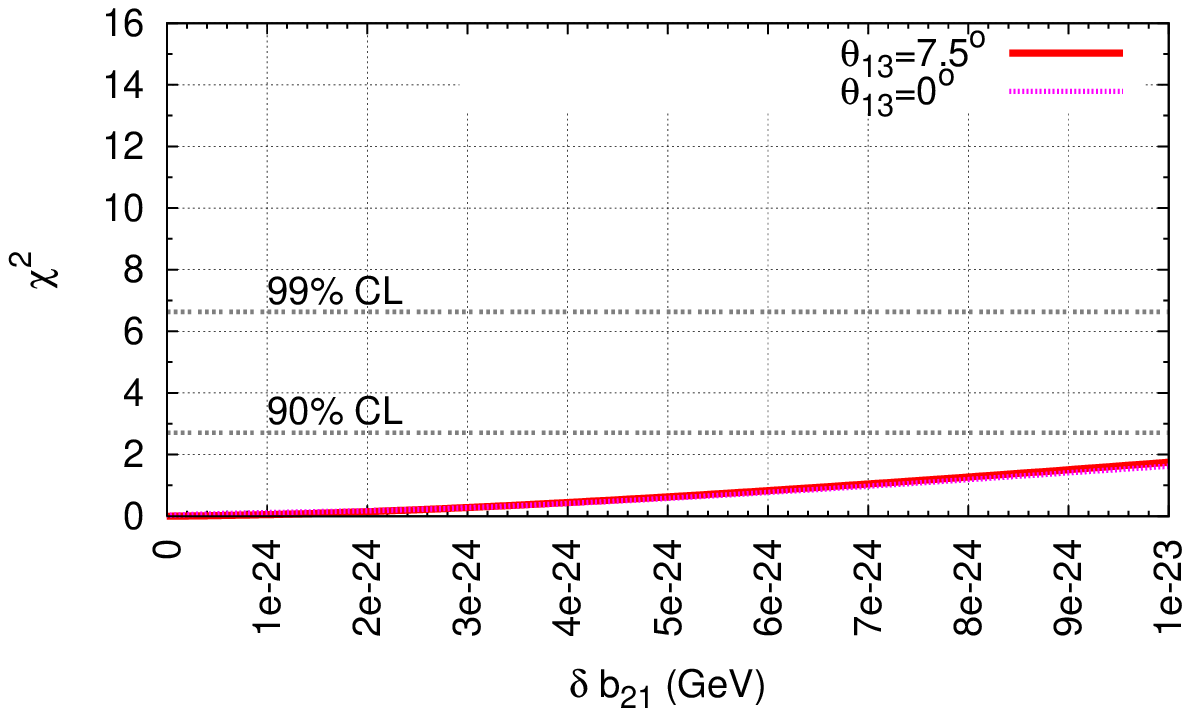}
\caption{\sf \small The bounds on $\delta b_{32}$ (left)  and $\delta b_{21}$ (right) 
for input $\theta_{13}=0^\circ$ and $7.5^\circ$. The curves for $\delta b_{21}$ with 
$\theta_{13}=0^\circ$ and $7.5^\circ$ have overlapped on each other.
We have set other oscillation parameters at their best-fit values and considered inverted hierarchy as input.
The marginalization has been carried out over whole allowed ranges of all oscillation parameters
(discussed in the text).
}
\label{f:bchi}
\end{figure*}

We have studied the atmospheric neutrino oscillation for 1 Mton.year exposure
(which is 10 years run of 100 kTon) of ICAL. The bounds on $\delta b_{21}$ and $\delta b_{32}$ are
presented in Fig. \ref{f:bchi}. It is found that the bounds are stronger for $\theta_{13}=0$
than its nonzero values. This can be understood in the following way. 
The atmospheric neutrinos covers  wide ranges of $E$ and $L$. 
The matter effect only plays role for neutrino with normal hierarchy and for
antineutrinos with inverted hierarchy. 
The matter resonance occurs in a limited zone of $E-L$ plane when $\theta_{13}$ is nonzero.
The resonance zones squeeze rapidly with decrease in $\theta_{13}$ values,
while the $CPTV$ contribution is independent of energy, baseline and $\theta_{13}$ values.
So, 
the $CPTV$ effect is only smeared out to some extent due to matter 
effect and bounds are weaker. It is very important to note that 
no `fake' $CPTV$ signal is arisen with atmospheric neutrinos, which 
may be the possibility in case of baseline experiments.
We have shown that a strong bound of $\delta b_{32} \lapp 6\times 10^{-24}$ GeV 
at 99\% CL can be obtained at a magnetized ICAL detector at INO, 
which is much stronger,  almost
one order of magnitude higher than those estimated till now for future experiments.
However, the bound on $\delta b_{21}$ is very poor, which is expected to improve 
in an analysis with solar neutrino data.

It can be understood that $CP$ and $CPT$ violation effect can be very cleanly
discriminated in  atmospheric neutrino oscillation data. 
The effect of $\theta_{13}$ and $\delta_{CP}$ appears dominantly 
neither in atmospheric neutrino oscillation nor in solar neutrino 
oscillation, but appears as subleading in both cases. These are 
observable in range of $E \sim 1$ GeV for atmospheric neutrino, 
where solar and atmospheric neutrino oscillations couple.
See Fig. 1 and 2 in Ref. \cite{Samanta:2009hd}.
It has also been shown  in Fig. 7 in Ref. \cite{Samanta:2009hd}  that 
if  $E_\nu > 2$ GeV, the sensitivity to $CP$ phase is very 
negligible. The sensitivity to $CPTV$ comes from the sensitivity
of $\Delta m_{32}^2$ of the experiment. From our previous work \cite{Samanta:2008af},
it can be understood that sensitivity to $\Delta m_{32}^2$  
will not be lost significantly if one consider events 
with $E_\nu > 2$ GeV. So, $E_\nu = 2$ GeV can be considered
roughly the boundary line for $CP$ and  $CPT$ violation studies
with atmospheric neutrinos.

The nonstandard interactions may mimic  as a signature of $CPT$ violation. 
However, this can also be separated out to an extent using atmospheric 
neutrino oscillation at a magnetized detector. In case of nonstandard
interactions, its contribution comes through matter resonance which
occurs for some  particular zones in $L-E$ plane of neutrinos with normal
hierarchy and for antineutrinos with inverted hierarchy.  These zones
does not move significantly with the change of oscillation parameters
for their present ranges of the uncertainties. On the
other hand, $CPTV$ effect comes over whole $L-E$ plane. 
So, again separating the resonance zones, one can separate out 
the $CPT$ violation from nonstandard interactions.

The separation of $CP$ violation, $CPT$ violation and nonstandard
 interactions can be  done for atmospheric neutrinos since it covers 
 wide ranges of $E$ and $L$ 
and contains both neutrinos and antineutrinos. This is a major advantage
of atmospheric neutrino experiment with a magnetized detector
over neutrino beams.

\section{Conclusion}
We have  studied $CPT$ violation in neutrino oscillation in a full
three flavor framework with matter effect constructing  a new way 
to find the oscillation probability so that $CPT$ violating 
terms are incorporated in the oscillation probability formula without any approximation. 
We carried out this study with
atmospheric neutrinos for a magnetized iron calorimeter detector
considering the muons (directly measurable with high resolution)
of the charge current events. We have shown that a stringent bound of
$\delta b_{32} \lapp 6\times 10^{-24}$ GeV at 99\% CL can be obtained 
at a magnetized ICAL detector with 1 Mton.year exposure, which is much stronger, almost 
one order of magnitude higher
than those estimated till now for future detectors. 
The advantages of atmospheric neutrinos to discriminate the effect of
$CPT$ violation, $CP$ violation and nonstandard interactions are also discussed.

{\it Acknowledgements:}
The author thanks Amol Dighe and Sandip Pakvasa for their comments and discussions
on the manuscript. 
This research has been supported by the  Neutrino Physics projects
at HRI. The use of excellent cluster computational
facility installed by the funds of this project  is also gratefully acknowledged.
The  general cluster facility of HRI has also been used for a part of this work.

\end{document}